\documentclass[12pt,english,showpacs,floats,amsmath,amssymb,nofootinbib]{iopart}

\usepackage{epsfig,epsf,amsthm,amsfonts,
            graphics,graphicx,amssymb,babel}

\def\centeron#1#2{{\setbox0=\hbox{#1}\setbox1=\hbox{#2}\ifdim
\wd1>\wd0\kern.5\wd1\kern-.5\wd0\fi
\copy0\kern-.5\wd0\kern-.5\wd1\copy1\ifdim\wd0>\wd1
\kern.5\wd0\kern-.5\wd1\fi}}
\def\ltap{\;\centeron{\raise.35ex\hbox{$<$}}{\lower.65ex\hbox{$\sim$}}\;}
\def\gtap{\;\centeron{\raise.35ex\hbox{$>$}}{\lower.65ex\hbox{$\sim$}}\;}

\newcommand{\Dzero}{D\O\ }

\def\met{\mbox{${\hbox{$E$\kern-0.6em\lower-.1ex\hbox{/}}}_T$}} 
\def\men{\mbox{${\hbox{$E$\kern-0.6em\lower-.1ex\hbox{/}}}_T$}} 

\newcommand{\TeV}{\ensuremath{\mathrm{Te\kern -0.1em V}}}
\newcommand{\GeV}{\ensuremath{\mathrm{Ge\kern -0.1em V}}}
\newcommand{\MeV}{\ensuremath{\mathrm{Me\kern -0.1em V}}}
\newcommand{\GeVcsq}{\ensuremath{\GeV\!/c^2}}

\newcommand{\fbinv} {fb$^{-1}$}

\newcommand{\mo}{m_{0}}

\newcommand{\miz}{m_{1/2}}
\newcommand{\tb}{\tan\beta}
\newcommand{\nall}{\tilde{{\chi}}^{0}}
\newcommand{\call}{\tilde{{\chi}}^\pm}
\newcommand{\ntwo}{\tilde{{\chi}}^{0}_{2}}
\newcommand{\cone}{\tilde{{\chi}}^\pm_{1}}
\newcommand{\none}{\tilde{{\chi}}^{0}_{1}}
\newcommand{\slep}{\tilde{l}^{\pm}}
\newcommand{\stau}{\tilde{\tau}}
\newcommand{\sneu}{\tilde{\nu}}

\renewcommand{\thetable}{\Roman{table}}
\addto{\captionsenglish}{
  \renewcommand{\figurename}{{\textsc{Fig.}}}
  \renewcommand{\tablename}{{\textsc{Table}}}
}

\newcommand{\beq}{\begin{equation}}
\newcommand{\eq}{\end{equation}}
\newcommand{\eeq}{\end{equation}}

\begin{document}
\lefthyphenmin=2
\righthyphenmin=3
%
\title[Addressing the multi-channel inverse problem]{
Addressing the multi-channel inverse problem at high energy colliders: 
a model independent approach to the search for new physics with trileptons\\
}
\author{S Dube$^1$, J Glatzer$^2$, S Somalwar$^3$, A Sood$^4$ and S Thomas$^3$}
\address{$^1$ Lawrence Berkeley National Laboratory, Berkeley, USA}
\address{$^2$ Fakult\"{a}t f\"{u}r Mathematik und Physik, Albert-Ludwigs-Universit\"{a}t, Freiburg i.Br., Germany}
\address{$^3$ Rutgers The State University of New Jersey, New Brunswick, USA}
\address{$^4$ University of California, Berkeley, USA}
\ead{sdube@lbl.gov}
%
%
\begin{abstract}
We describe a method for interpreting trilepton searches at
high energy colliders in a model-independent fashion and apply it to
the recent searches at the Tevatron.  The key step is to
recognize that the trilepton signature is comprised of four experimentally
very different channels defined by the number of $\tau$ leptons in the trilepton
state. Contributions from these multiple channels to the overall
experimental sensitivity (cross section times branching ratio) are
model-independent and can be parametrized in terms of relevant new
particle masses.  Given the trileptonic branching ratios of a specific model,
these experimentally obtained multichannel sensitivities can be combined
to obtain a cross section measurement that can be used to confront 
the model with data. Our model-independent results are more widely applicable than 
the current Tevatron trilepton results which are stated exclusively in terms
of mSUGRA parameters of supersymmetry.  The technique presented here
can be expanded beyond trilepton searches to the more general ``inverse
problem'' of experimentally discriminating between competing models
that seek to explain new physics discovered in multiple channels.
\end{abstract}
\submitto{\JPG}
\maketitle

\section{Introduction}

The canonical method to search for signatures of new physics at high
energy colliders is through the use of high transverse momentum
objects such as jets, isolated photons, electrons and muons, often accompanied by
transverse momentum imbalance.
Isolated electrons and muons are easier to identify and suffer
smaller backgrounds than jets. A classic new physics search that makes
use of these relatively clean objects is inclusive trileptons plus
missing transverse energy.

Standard Model background for trileptons can be mostly
estimated using data-driven techniques with minimal
reliance on monte carlo simulations. This signature also covers a 
wide range of new physics scenarios.

\begin{figure}
  \centering\epsfig{file=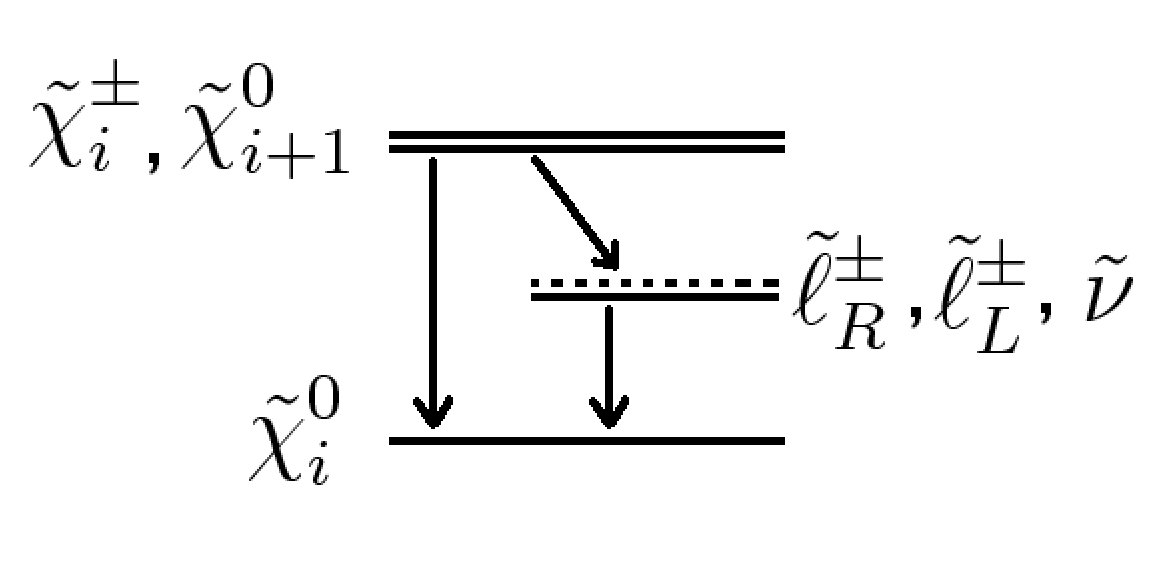,height=5cm}
  \caption{
Supersymmetric cascade transitions relevant to the trilepton plus missing
transverse energy signature. The decay of the $\nall_{i+1}$ and $\call_{i}$ can proceed
via on-shell or off-shell gauge bosons ($W,Z$) or via intermediate supersymmetric states ($\slep$, $\sneu$).
}
  \label{fig:tril}
\end{figure}

The most widely discussed possibility for new physics that could give
rise to the trilepton signature is supersymmetry
(SUSY)~\cite{SUSYref}. In this case trileptons can arise from 
chargino-neutralino production followed by their cascade decays.
The chargino, $\cone$, decays to the lowest neutralino, $\none$, which is the 
stable lightest supersymmetric particle (LSP), yielding a
charged lepton and neutrino, while the 
neutralino, $\ntwo$, yields two charged leptons and the LSP. The stable neutralinos
and neutrino escape undetected and thus carry away
transverse energy. This gives the three leptons with
missing transverse energy ``trilepton'' signature~\cite{oldtrilepton}, 
as depicted schematically in Figure~\ref{fig:tril}.

Although the trilepton signature can arise over a fairly wide region
of SUSY parameter space, it is often discussed within the context of
specific model assumptions or special subspaces of the general
parameter space. The most widely abused subspace in this regard is one
with a universal mass for all scalars, $m_0$, and another for all
gauginos, $m_{1/2}$, both defined at 
the unification scale. For obscure historical reasons, this SUSY
subspace is referred to as minimal supergravity (mSUGRA).

The Tevatron trilepton searches have traditionally been
interpreted within the context of mSUGRA parameters, or
experimentally convenient variations thereof. Both CDF~\cite{cdf2fb} 
and \Dzero~\cite{d0trilep,d02fb} 
collaborations show results as a function of 
$m_0$ and $m_{1/2}$, while holding other parameters fixed
(Section~\ref{sec:review}). 
Experimentalists resort to mSUGRA for
conveying trilepton results not because of its theoretical
merit, but for the practical reason that it
has fewer parameters than other choices. Even then, not all
of its five parameters are covered.
This model-specific 
restrictive approach makes it difficult to interpret search results for more
general superpartner mass spectra and mixings or for other new physics
models besides supersymmetry.  

Below, we describe an economical way to state the trilepton results in a
model-independent fashion. Section~\ref{sec:threepr} lays out
three main principles that allow experimentalists to distance
themselves from specific models while interpreting search
results. In there, we also identify the multiple channels 
that constitute the trilepton signature
and formulate a scheme for parametrizing their experimental
sensitivities as a function of relevant phenomenological parameters. 

Section~\ref{sec:meth} has the complete scheme for presenting
search sensitivities in a model-independent fashion. Also
included is the derivation of an equation for combining these sensitivities 
using the particle masses
and branching ratios predicted by a model under consideration
to obtain the experimental cross section measurement specifically for that
model. If this measured value agrees with the cross section predicted
by the model, the model is consistent with the data.

In section~\ref{sec:sigmaB}, we take on the task of converting the Tevatron
trilepton results into a model-independent format. We formulate a 
suitable scheme for
parametrizing the trilepton multichannel sensitivities as a function of the mass
parameters and then evaluate the coefficients of parametrization using
CDF results in section~\ref{sec:simcdf}. Section~\ref{sec:fitres} has
the resultant model-independent formulation of the CDF trilepton
results and a simple example to show how to use these results. In the
following sections~\ref{sec:recdf}~and~\ref{sec:otherm}, we use these
results to recover CDF's mSUGRA trilepton interpretation, to
project sensitivities to a significantly larger data sample,
and to address a couple of non-mSUGRA scenarios.

Finally, in section~\ref{sec:invprob}, we briefly describe how the
technique we use here for trileptons may be used more generally in
the context of deciphering the nature of new physics that manifests 
itself in multiple channels.

The model-independent experimental 
parametrizations presented in this article are electronically
available in the form of a spreadsheet utility
at Rutgers University's Department of Physics publication
archive website~\cite{pubarch}.


\section{A Brief Review of CDF and \Dzero Trilepton Searches}
\label{sec:review}

At the Tevatron, CDF~\cite{cdf2fb} and \Dzero~\cite{d0trilep,d02fb} have
conducted similar trilepton analyses using around 2-3~\fbinv~data
each. The final state consists of at least two electrons or muons, and
both experiments allow an isolated track in lieu of the third lepton 
in order to add sensitivity for the $\tau$-lepton decays.

\begin{itemize}
\item {\bf CDF}: CDF interpretes its results in the context of the
canonical mSUGRA scenario which has five parameters
and considerably simplifies the large MSSM parameter space. CDF
conducts an exclusive multichannel analysis~\cite{thesis}
where events are sorted based on the
expected signal purity, and the results combined in the end. The
analysis is split into channels with three electrons and/or muons, and
those where the third object is an isolated track. The published cross section
results, presented for 2~\fbinv~of data, are shown as a function of
two of the mSUGRA parameters ($\mo$, $\miz$) or as a function of the
$\cone$ mass while keeping the other mSUGRA parameters fixed at some
suitable values (tan($\beta$)=3, $A_0$=0 and $\mu>0$). This
intractable choice
immediately begs the question of search
sensitivity for other models as well as for other values of the mSUGRA 
parameters.

\item{\bf \Dzero}: \Dzero interpretes its results in the context of
the mSUGRA scenario, as well as an MSSM scenario that follows all the
mSUGRA mass constraints. The \Dzero analysis has channels
with two electrons and/or muons where the third object is an isolated
track, and a channel with one electron or muon, one hadronically
reconstructed tau-lepton, and one isolated track. The final results
are presented in terms of the mSUGRA parameters $\mo$, $\miz$ and
$\tb$. In the same way as CDF, the \Dzero results cannot be
reinterpreted easily when the theory parameters are different from
those assumed in the standard result.

\end{itemize}

\section{Three Organizing Principles}
\label{sec:threepr}

There are three major hurdles in stating trilepton results in a
model-independent fashion. First, there are several channels that make
up the trilepton signature. Second, it is not clear exactly what
experimental information to give out as a measure of experimental
sensitivity since these channels have different acceptances and
Standard Model backgrounds. Finally, the experimental acceptance
depends on the nature of signal in a phenomenological model of
interest. 

A brute force approach of publishing experimental acceptance
and background on a channel-by-channel basis for different models is
not only prohibitively cumbersome, but would also require the reader
to delve deeply into experimental details.  However, volumes of data 
need not be published to achieve model-independence if experimental
sensitivity can be quantified concisely and stated separately for a small set of
phenomenologically important channels and parameters of universal
interest. We now discuss these three requirements one by one.

\subsection{Experimental Sensitivity $\{\sigma \! B\}$}
\label{sec:expsens}

The reach of an experimental search (or measurement) depends on the
amount of collected data (integrated luminosity), detector's
acceptance for the signal, and the extent of Standard Model
background.  A concise and commonly used experimentally accessible
quantity that characterizes the overall search sensitivity is the
product of production cross-section and the branching ratio,
$\{\sigma \! B\}$. To measure it, the number of standard model background is
subtracted from the observed number of events and then the detector
acceptance and the integrated luminosity are divided out.
$\{\sigma \! B\}$ is
enhanced by optimizing the selection criteria (``cuts'') to increase
acceptance while keeping the backgrounds in check.  
A model is successfully confronted by the experiment if the measured
$\{\sigma \! B\}$ is comparable to the model's value for the signal.  
If an experiment
fails to find the signal, the $\{\sigma \! B\}$ sensitivity is typically
expressed as a 95\% confidence level upper limit.

Note that $\{\sigma \! B\}$ subsumes the knowledge of
detector acceptance, backgrounds and integrated luminosity; it
thus serves as the sole indicator of the experimental reach. 
Since the detector acceptance depends on the nature of the signal,
$\{\sigma \! B\}$ is a model-dependent quantity. If it is to
be used in a model-independent context, a way must be
found to measure and tabulate it as generically as possible so that
it can be used to reconstruct the sensitivity for 
other models.

\subsection{Identifying Relevant Multichannels}
\label{sec:Relmult}

Experimental search is often carried out in multiple channels to cover
as much signal as possible. As described above, CDF devotes an analysis
channel for its highest quality electrons and muons, and a separate
one for leptons that are not as well reconstructed. Further, in order
to include the short-lived $\tau$ lepton as one of the trileptons,
there is another higher-background trilepton channel with two leptons
and an isolated track which serves as a proxy for the $\tau$ lepton.
These channels suffer from different amount of Standard Model
backgrounds and have varying detector acceptance. 

These experimental search channels described in terms of electron and
muon quality or the presence of a track do not
have direct relevance from a phenomenological point of view. However,
their overall ability to detect the $\tau$ lepton is of interest
because the $\tau$ flavor content of the trilepton state is an
important clue to the nature of new physics. Another important reason
to focus on the $\tau$ lepton is that the experimental acceptance
depends drastically on the number of $\tau$ leptons in the trilepton state
because the detection of hadronic decays of the $\tau$ lepton draws a
substantial background.  Detecting trileptons when none of the leptons
are $\tau$'s is straightforward and detecting one $\tau$ via its
hadronic decays is manageable. However, detection of trileptons with 
two (three)
$\tau$'s requires that at least one (two) $\tau$ leptons decay
leptonically.  Since the leptonic decay of the $\tau$ lepton takes
place approximately one-third of the time, even a trilepton search
with only electrons and muons in its search channels will still
have indirect sensitivity to the three physics channels above that
contain $\tau$ leptons.

The presence of $\tau$ leptons in the trilepton final state greatly
affects the search sensitivity and is thus a major source of
model-dependence.  
Therefore, the combined $\{\sigma \! B\}$
sensitivity gleaned from the multiple experimental channels should be
mapped onto $\{\sigma \! B\}$ sensitivities tabulated in terms of the $\tau$
content of the signal. 
Accordingly, we classify the trilepton signature into four exclusive channels based
on their $\tau$-content:
\begin{itemize}
\item 0$\tau$
\item 1$\tau$ (In SUSY language, e.g., because $\cone \rightarrow \tau\nu\none$).
\item 2$\tau$ ($\ntwo \rightarrow \tau\tau\none$).
\item 3$\tau$ (Both).
\end{itemize}

We will denote the measured experimental sensitivity for a channel
that contains $i~\tau$ leptons by $\{\sigma \! B\}_i$.  Each $\{\sigma \! B\}_i$
is measured by assuming a 100\% branching fraction for
the channel with $i~\tau$ leptons, i.e., that the trilepton signal
contains exactly $i~\tau$ leptons. Each of the four $\{\sigma \! B\}_i$'s
receives varied contributions from the underlying experimental
search channels which are characterized by how many electrons, muons
and tracks they contain, how well reconstructed these objects are,
and so on.

\subsection{Identifying Universal Parameters}
\label{sec:Unipar}

The final principle in achieving model-independence is to avoid the
parameters of specific models and express the results as a function of
the underlying parameters such as particle masses and mass differences that
characterize the signature decay.  In case of trileptons and
supersymmetry, the chargino, neutralino, and slepton masses control
the production and cascade decays responsible for the trilepton
signature.  Their masses and mass differences are the appropriate
parameters for expressing the experimental results as opposed to the
$m_0$ and $m_{1/2}$ parameters of mSUGRA. Therefore, the 
$\{\sigma \! B\}$ search sensitivity can be expressed in a
model-independent fashion if it is given as a function
of these universal mass parameters.

For trileptons, we pick
three mass parameters for this purpose.  The scale for missing energy is set
by the mass $M$ of the undetected new particle (lower state). The mass
difference $\Delta M_1$ between the upper state and a possible
intermediate state
plays a role in the distribution of trilepton
momenta. Finally, the total energy available to the decay
products is given by $\Delta M_2$, the mass difference between
the upper and the lower state. For the specific instance of
trileptons from supersymmetric chargino-neutralino decays, $M$ is 
the LSP mass, $\Delta M_1$ is the mass difference between the 
chargino and the intermediate right-handed slepton and 
$\Delta M_2$ is the chargino-LSP mass difference.

In supersymmetric theories that have direct production of 
Wino-like states, the mass of the 
lightest chargino is approximately equal to one of the neutralinos.
We have thus made a concession to the supersymmetric
origin of the trilepton signature in assuming 
that the two upper states have the same mass.
We also allow at most one intermediate state particle.
The most general treatment would require additional mass parameters beyond the
three we employ.

\section{A Recipe for Model-independent Interpretation}
\label{sec:meth}

At this point, we claim that model-independence can be achieved, i.e.,
the trilepton search results can be used for confronting an arbitrary model that
predicts trilepton signal, provided (a) the search results are tabulated
in terms of separate $\{\sigma \! B\}_i$ measurements for the four
trilepton $\tau$ subchannels, and (b) the $\{\sigma \! B\}_i$'s are stated as a
function of the mass parameters described above.
We substantiate this claim by showing how to deduce
the experimental cross section measurement for a model under consideration, 
using the masses and the four $\tau$ branching ratios predicted by the model.

Consider a model with cross section $\sigma$ for producing the parent
states such as chargino-neutralino that lead to the trilepton
signature. The model also predicts branching ratio $B_i$'s for trilepton channels with
$i~\tau$'s. On the experimental side, let us assume the detector
acceptance (efficiency) for these channels to be $A_i$ and the collected luminosity
to be $L$. Then the total number of signal trilepton events observed by the experiment, $N$, is 
given by $N = \sum_{i=0}^3 L \sigma B_iA_i$. Rearranging this we get $1/\sigma = \sum_{i=0}^3 (LA_i/N)B_i$.

Now recall from above that the experimental sensitivities  $\{\sigma \! B\}_i$ were calculated 
four times (independently) by pretending that in turn that all of the very 
same N observed events came entirely (i.e. with 100\% branching ratio) from signals 
with 0,1,2 and 3 tau’s.
Therefore, it also follows that 
$N = L\{\sigma \! B\}_i A_i$. Equating the two $N$'s gives the desired equation
to calculate the experimental cross section measurement for a model:
\begin{equation}
\frac{1}{\sigma_{XM}} = \displaystyle\sum_{i=0}^3 \frac {B_i}{\{\sigma \! B\}_i}.
\label{eq:sigmaB}
\end{equation}
We have added a subscript $XM$ on the cross section to indicate that 
$\sigma_{XM}$ is the cross section as measured by the experiment for the model. Note
that the equation is free of experimental details such as acceptance, number of events 
and luminosity. $\sigma_{XM}$ is an aggregate of the $\{\sigma \! B\}_i$ multichannel cross section 
measurements.

To recapitulate, the experiment provides its multichannel
$\{\sigma \! B\}_i$ search sensitivities for each of the four trilepton
$\tau$ channels as a function of the three mass parameters.  The model
provides the (three) relevant sparticle mass parameters and the (four)
multichannel trilepton branching ratios, $B_i$'s.
Equation~\ref{eq:sigmaB} then blends the phenomenological
and experimental information to give the experimental
reach, quantified as cross section for the model under
consideration, $\sigma_{XM}$. The model is confronted experimentally if
the trilepton production cross section it predicts exceeds $\sigma_{XM}$.
We reiterate that the experiment need not
separately provide the detector acceptance or the Standard Model background 
information
as this information is already incorporated in the $\{\sigma \! B\}_i$ sensitivity
measurements. Also note that $\sigma_{XM}$ is not a pure
experimental quantity since models with differing trilepton
branching ratios would yield different $\sigma_{XM}$'s from 
the same experimental data.


\section{$\{\sigma \! B\}_i$ Search Sensitivities as Function of Generic Mass Parameters}
\label{sec:sigmaB}

As described above, the three mass parameters in our scheme are 
$M$, $\Delta M_1$, and $\Delta M_2$. We obtain acceptances $A_i$ as described in section~\ref{sec:simcdf}
along with the background estimate from Ref.~\cite{cdf2fb} to calculate the sensitivities
as a function of the three mass parameters.
We empirically find that the
dependence of $\{\sigma \! B\}_i$ on $M$ conveniently factors out from the 
$\Delta M_1$ and $\Delta M_2$ dependence. This is somewhat expected since $M$
essentially determines the amount of missing energy in the final state, whereas
$\Delta M_1$ and $\Delta M_2$ determine the amount of energy available for the leptons. Thus, we have
\begin{equation}
\{\sigma \! B\}_i^{-1} = f_i(M) \times h_i(\Delta M_1,\Delta M_2)
\label{eq:2bd_1}
\end{equation}
with $i=0,1,2,3$ indicating the $\tau$ content and $f_i$ and $h_i$ are parametric functions to be
determined from data.

In the absence of an intermediate
particle (say when the $\slep$ is heavier than the $\cone$), we can disregard
$\Delta M_1$, giving
\begin{equation}
\{\sigma \! B\}_i^{-1} = f_i(M) \times g_i(\Delta M_2)
\label{eq:3bd_1}
\end{equation}
where $g_i$ is another parametric function.  These functions can be written as Taylor expansions: 
\begin{equation}
f(M) = 1 + a_1 (M) + a_2 (M)^2
\label{eq:fnew}
\end{equation}
\begin{equation}
g(\Delta M_2) = b_0 + b_1  (\Delta M_2) + b_2  (\Delta M_2)^2
\label{eq:gnew}
\end{equation}
\begin{eqnarray}
h(\Delta M_1, \Delta M_2) & = & c_0 + c_1  (\Delta M_2) + d_1  (\Delta M_1) \nonumber \\
& & + c_2  (\Delta M_2)^2 + d_2 (\Delta M_1)^2 \nonumber \\
& & + e_2  (\Delta M_1 \times  \Delta M_2)  .\label{eq:hnew}
\end{eqnarray}
Note that the subscripts $i$ denoting the $\tau$ content are implicit for the functions $f,g$ and $h$ on
the left hand side as well as for the coefficients on the right hand side of these equations. We have 
normalized $f$ to the sensitivity at $M=0$ ($f(0)=1$). This effectively makes $f$ a ``scaling'' function; $f(M)$
is a multiplicative factor to the $g$ and $h$ functions (which have units of picobarns).

The values of the expansion coefficients would be determined from
experimental data. The coefficients in turn define the $f,g$ and $h$
functions and thereby the $\{\sigma \! B\}_i$ sensitivities as a function
of the masses.  Since the Tevatron trilepton results are confined to
the mSUGRA scenario, we calculate these coefficients with the aid of
extensive standalone simulation and using public information from
the CDF collaboration.  We describe this procedure in the next section before giving the
results.

\section{Determination of $\{\sigma \! B\}_i$ Sensitivity Parametrization Using CDF Results}
\label{sec:simcdf}

We mimic CDF analysis using {\sc pythia} v6.409~\cite{pythia} to generate numerous
samples of simulated trilepton events for different choices of the
mass difference between $\cone$ and $\none$, and $\cone$ and
$\slep$. See Ref.~\cite{juls} for further details of sample generation.  
Subsamples for each of the four $\tau$ channels undergo identical analysis. 
We select events with
three electrons or muons, or events with two electrons or muons
and any other charged particle, with $p_{\rm T}$ thresholds similar to
CDF's. The charged particle selection catches the $\tau$ lepton
single-prong decays and it is required to be isolated\footnote{The sum
of $p_{\rm T}$'s of other charged particles within an $\eta-\phi$ cone
of 0.4 is required to be less than 10\% of the $p_{\rm T}$ of the
charged particle.} to approximate the isolated track selection used by
CDF. We calculate the missing energy $\men$ by taking the vector sum
of the transverse momenta of all neutrinos and LSP's present in the
event. 

\begin{table}[htbp]\centering
  \caption{ Selections for Pythia based simulations to mimic the Tevatron trilepton searches. CDF and \Dzero selection
criteria tend to be fairly similar.(OS=opposite-sign)}
 \begin{tabular}{cc}
\hline \hline
Variable & Selection \\ \hline
$p_{\rm T}^{1,2,3}$& $>$ 15, 5, 5~GeV/$c$ \\
$|\eta^{1,2,3}|$&  $<$ 1.1 \\
$\men $ & $> 20$ GeV \\
max OS Mass& $> 20~\GeVcsq$, $\notin [76,106]~\GeVcsq$ \\
next OS Mass & $> 13~\GeVcsq$, $\notin [76,106]~\GeVcsq$ \\
\hline \hline
 \end{tabular}
 \label{tab:pytsel}
\end{table}

Other selections listed in Table~\ref{tab:pytsel} are made and
the final $\{\sigma \! B\}_i$ sensitivities for
each subsample are calculated using the acceptance for each subsample and the total background
as measured by the CDF experiment~\cite{cdf2fb}. The selections
follow those of the CDF analysis described in Ref.~\cite{thesis} and
are meant to reproduce the CDF analysis.

\begin{figure}
\begin{minipage}{0.5\linewidth}
  \centering\epsfig{file=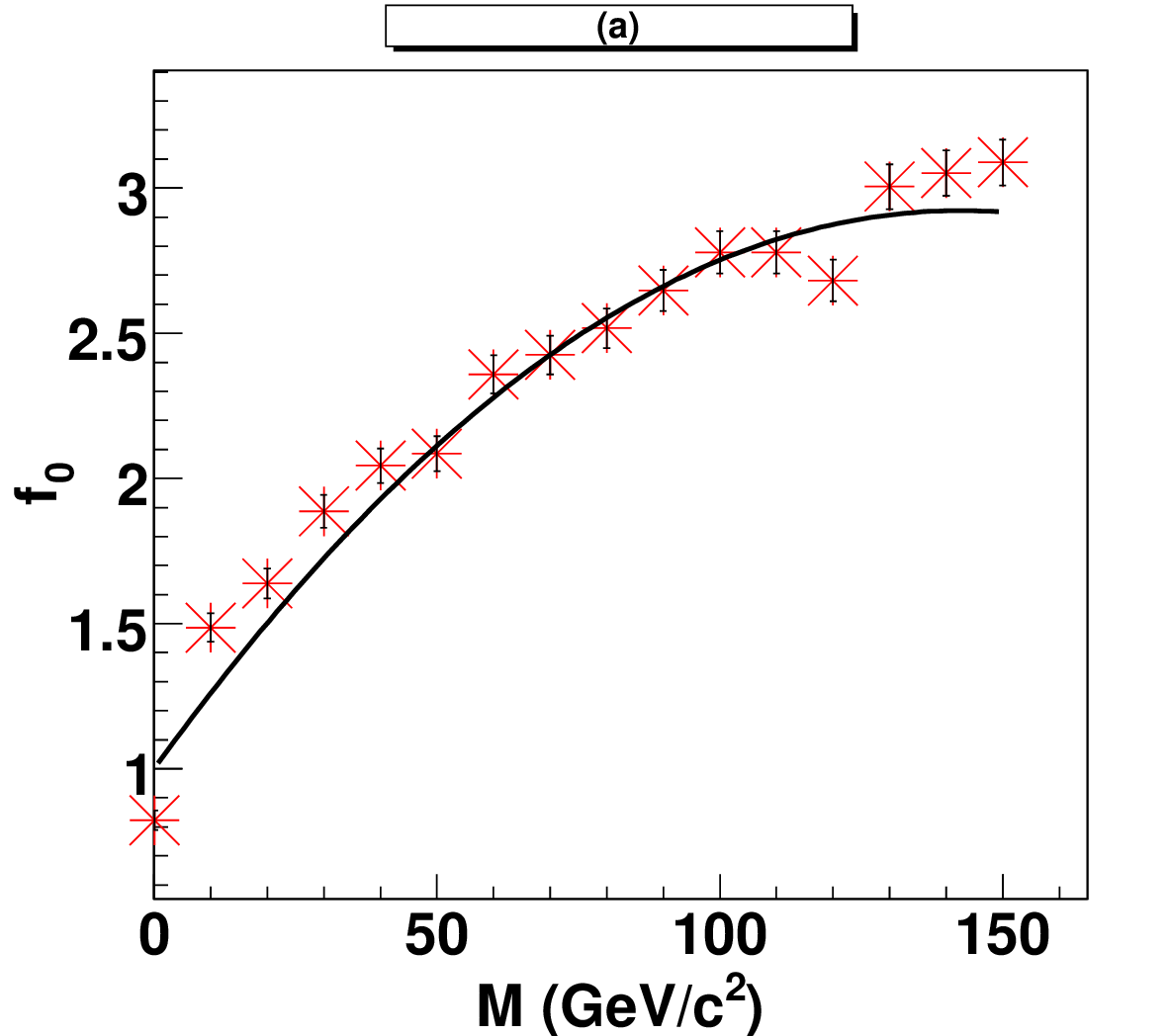,width=\linewidth}
\end{minipage}\hfill
\begin{minipage}{0.5\linewidth}
  \centering\epsfig{file=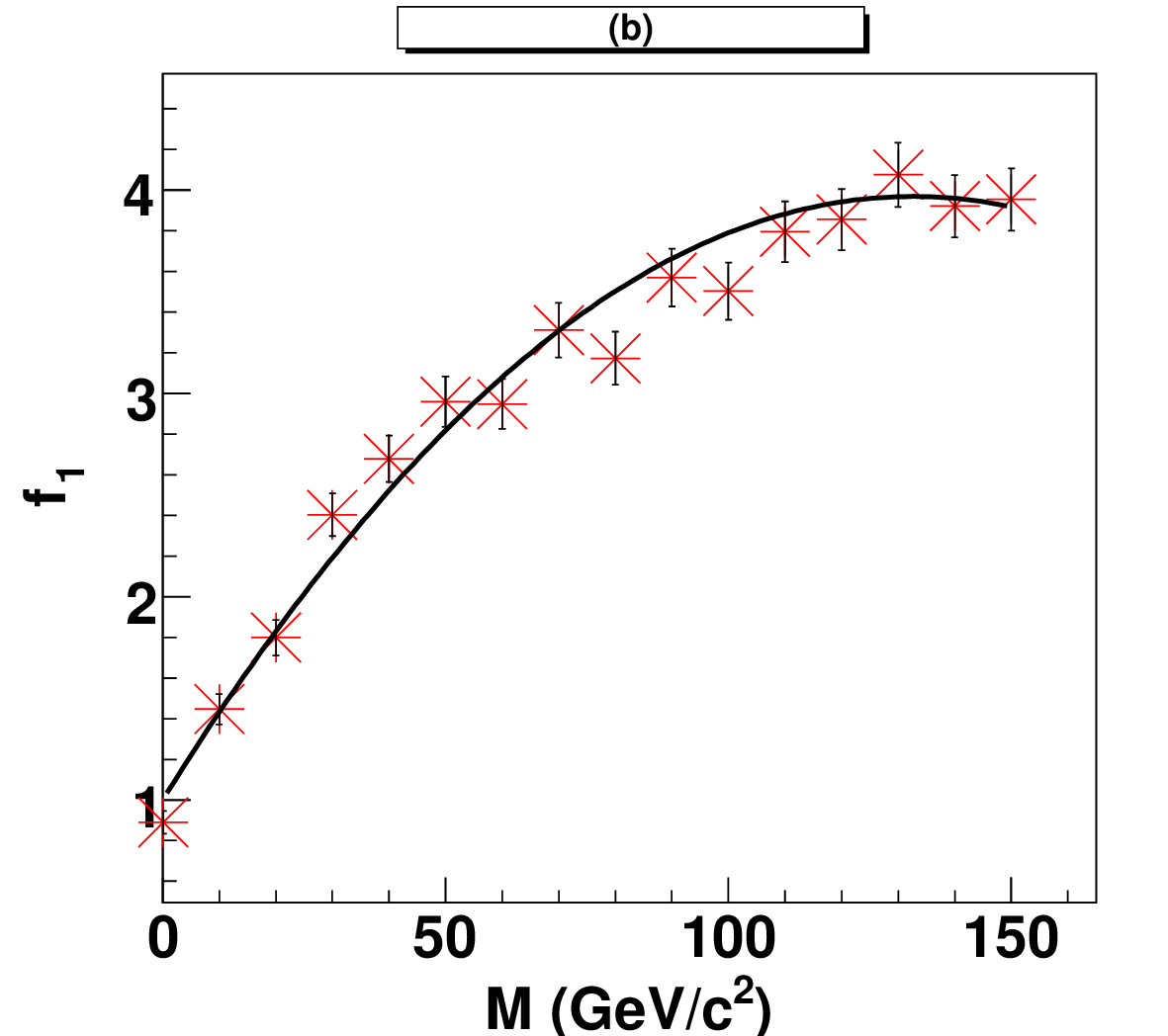,width=\linewidth}
\end{minipage}

\caption{The scaling function $f(M)$ is shown for the $0\tau$ (a) and $1\tau$ (b) subsamples.
The points are the actual measured values. The results of a fit using Eq.~\ref{eq:fnew} are shown (with coefficients
as given in Table~\ref{tab:par}). $\Delta M_1$, and $\Delta M_2$ are fixed at 25 and
  50~\GeVcsq, respectively. }
  \label{fig:lspfit1}
\end{figure}

We systematically scan the $\Delta M_1$-$\Delta M_2$ space
while maintaining the following mass relations :\\ M($\cone$) $=$
M($\ntwo$), M($\tilde{e}_R$) $=$ M($\tilde{\mu}_R$) $=$
M($\stau_1$). The decay of $\cone$ and $\ntwo$ to $\sneu$'s is turned
off and thus the $\sneu$'s play no further part in the analysis. We
set M($\none$) $=$ 70~\GeVcsq~and vary $\Delta M_2$ from 40 to
90~\GeVcsq~in steps of 5~\GeVcsq.

The two cases, positive $\Delta M_1$, and negative $\Delta M_1$ are
treated separately. In the case of negative $\Delta M_1$, 
the $\{\sigma \! B\}_i$ sensitivities
depend only on $\Delta M_2$. When $\Delta
M_1$ is positive, i.e., there is an intermediate state, 
we vary its mass M($\slep$) from a value 5~\GeVcsq~higher than M($\none$), to
5~\GeVcsq~less than M($\cone$) in steps of 5~\GeVcsq.
In all cases, the events are split
into four subsamples with 0, 1, 2 and 3$\tau$'s
in the final state.

We then parametrize the $\{\sigma \! B\}_i$ sensitivities 
(for each subsample) as a function of $\Delta M_1$, $\Delta M_2$ and $M$. 
First we determine the dependence on the overall mass scale by fixing the relative masses of M($\cone$), M($\slep$) and
M($\none$). We fix $\Delta M_1 = 25$~\GeVcsq, and $\Delta M_2 = 50$~\GeVcsq~and
determine $f(M$) (normalizing it to $f(0)$). This is shown in Figure~\ref{fig:lspfit1} for the 0$\tau$ and 1$\tau$ cases. 
The fit parameters are given in Table~\ref{tab:par}.

We now proceed to determine the term which depends on the mass differences between M($\cone$), M($\slep$) and M($\none$).
We fix M($\none$) at 70~\GeVcsq, and measure the 
$\{\sigma \! B\}_i$ sensitivities 
as a function of \\
a) $\Delta M_2$, if M($\slep$)$>$M($\cone$), and given by  $g(\Delta M_2)$, or
b) $\Delta M_1$ and $\Delta M_2$, if M($\slep$)$<$M($\cone$), and given by $h(\Delta M_1, \Delta M_2)$.

\begin{figure} [htb]
  \centering\epsfig{file=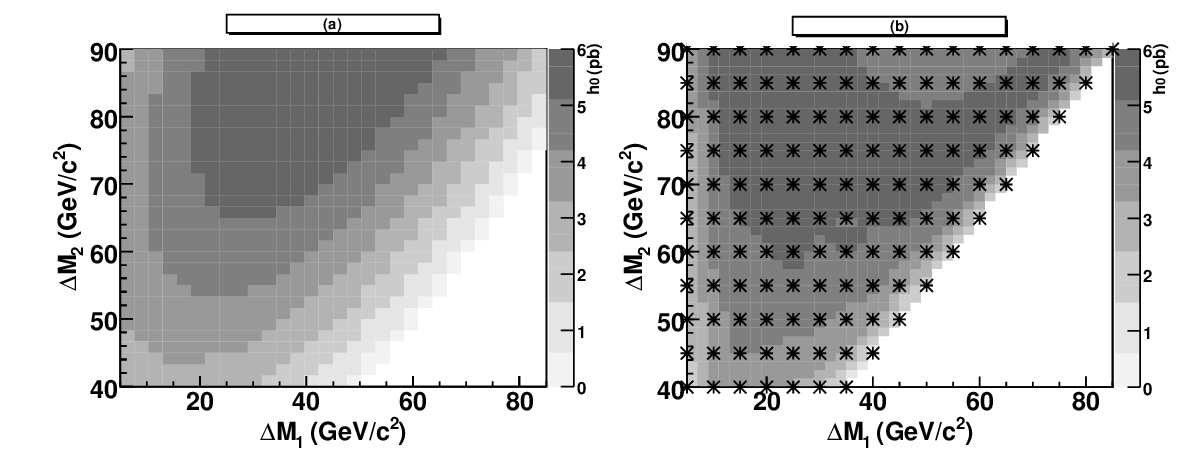,width=\linewidth}
  \caption{ The measured values of function $h$ are shown for the $0\tau$ case in 
(b) for the marked points along with the extrapolation using ROOT~\cite{root}.
The result of a fit using Eq.~\ref{eq:hnew} with parameters given in Table~\ref{tab:par} is shown in (a). The parameter
$M$ is fixed at 70~\GeVcsq.}
  \label{fig:tau0fit}
\end{figure}

\begin{figure} [htb]
  \centering\epsfig{file=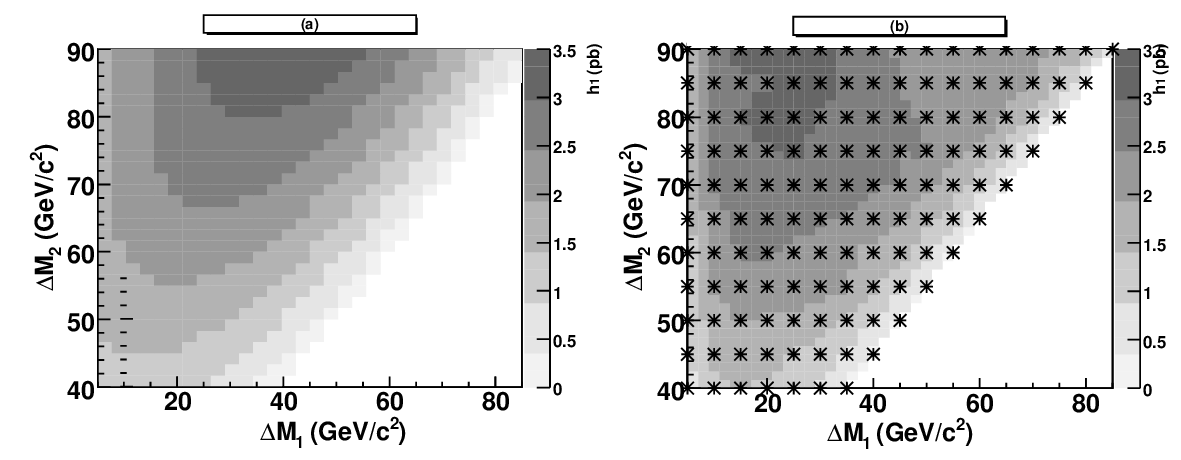,width=\linewidth}

  \caption{ The measured values of function $h$ are shown for the $1\tau$ case in 
(b) for the marked points along with the extrapolation using ROOT~\cite{root}.
The result of a fit using Eq.~\ref{eq:hnew} with parameters given in Table~\ref{tab:par} is shown in (a). The parameter
$M$ is fixed at 70~\GeVcsq.}
  \label{fig:tau1fit}
\end{figure}

\begin{figure}
\begin{minipage}{0.5\linewidth}
  \centering\epsfig{file=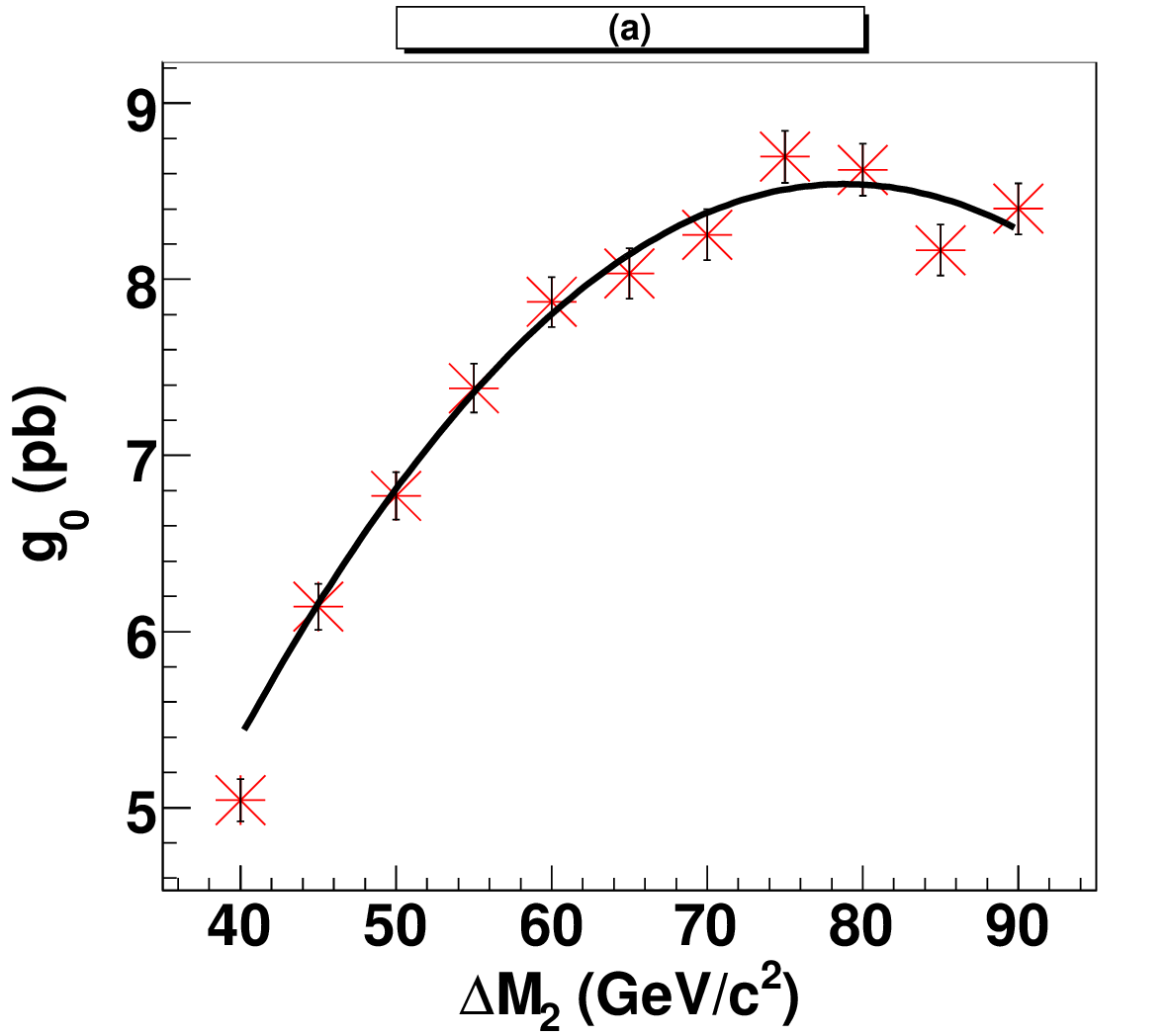,width=\linewidth}
\end{minipage}\hfill
\begin{minipage}{0.5\linewidth}
  \centering\epsfig{file=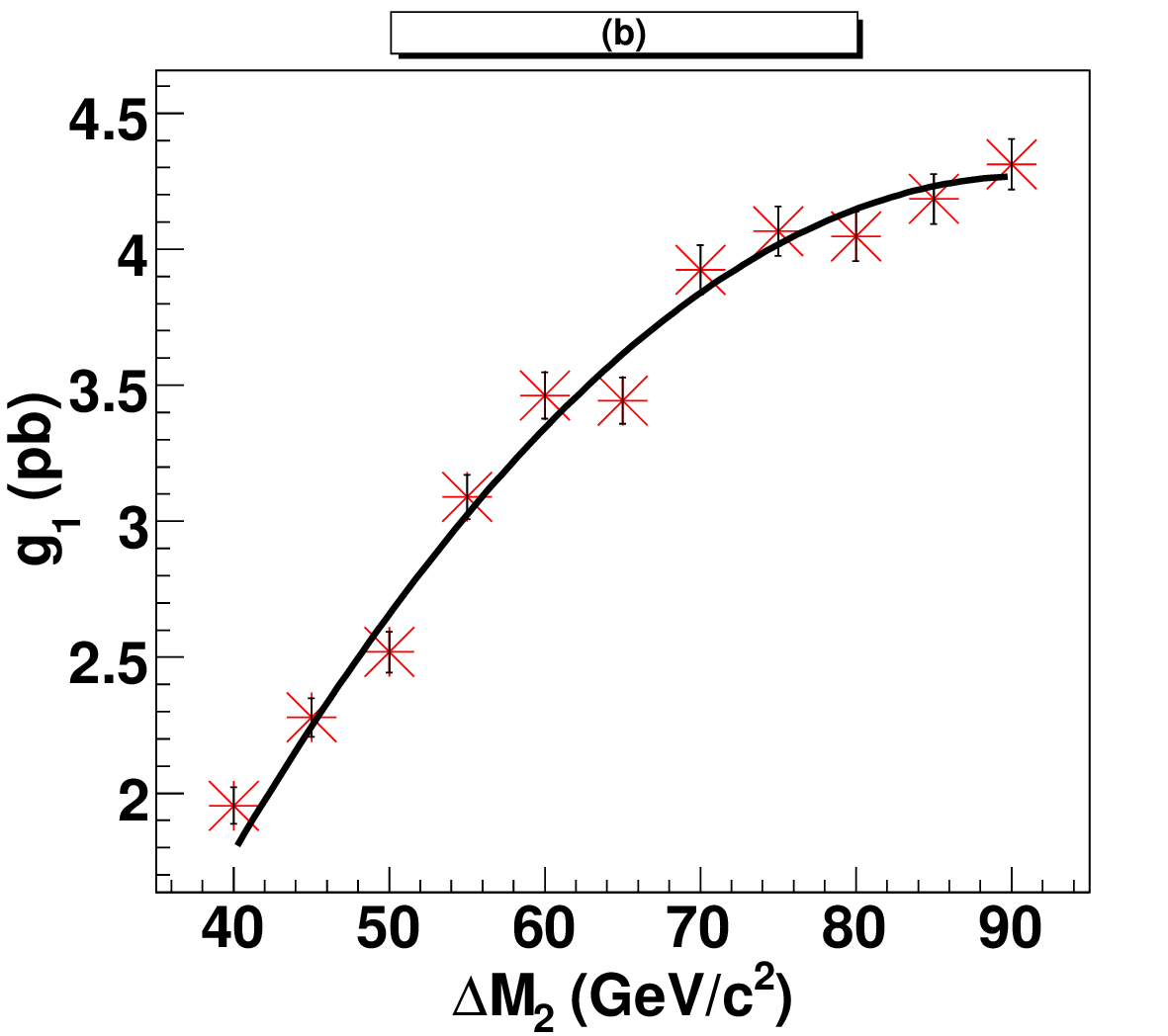,width=\linewidth}
\end{minipage}

\caption{The function $g(\Delta M_2)$ is shown for the $0\tau$ (a) and $1\tau$ (b) subsamples.
The points are the actual measured values. The results of a fit using Eq.~\ref{eq:gnew} are shown (with coefficients
as given in Table~\ref{tab:par}). The parameter $M$ is fixed at 70~\GeVcsq. }

  \label{fig:dmfit2}
\end{figure}

The actual $g(\Delta M_2)$ and $h(\Delta M_1, \Delta M_2)$ are obtained by dividing out $f(M=70)$ from 
the measured $\{\sigma \! B\}_i$. Figures~\ref{fig:tau0fit} and \ref{fig:tau1fit} show the measured and fitted 
$h$ for the $0\tau$ and $1\tau$ case respectively. Figure~\ref{fig:dmfit2} shows the measured $g$ for
the same cases.

The final
$\{\sigma \! B\}_i$ sensitivities are then calculated by multiplying the $f$ and $g$ (or $h$) functions.

\section{Model-Independent Trilepton Results}
\label{sec:fitres}

The values of the parameters for the $f,g$ and $h$ functions that
determine the $\{\sigma \! B\}_i$ sensitivities 
(See Eqns.~\ref{eq:3bd_1}-\ref{eq:hnew}) obtained
from our analysis of the CDF 2~fb$^{-1}$ trilepton search are
given in Table~\ref{tab:par}.  Several comments are in order. These
results are obtained with 2~fb$^{-1}$ integrated luminosity.  To apply
them to different amount of data requires scaling by the
square-root of the luminosity. Also note that these results describe a
null finding in terms of 95\% confidence level upper limits. When
searching for new physics, it is customary to express the sensitivites
in units of standard deviations, e.g., a 3~$\sigma$ or 5~$\sigma$ discovery
potential. (95\% C.L. upper limit sensitivity amounts to a 1.64~$\sigma$
measurement sensitivity.) Finally, we do not explicitly write the units of the
coefficients in the table. They are such that when the 
mass parameters are in
units of GeV/c$^2$, the resulting cross-section values are in picobarns.

The parametrization is valid when the lower state mass $M$
is less than 150~\GeVcsq~and the upper-lower state mass difference
$\Delta M_2$ is less than 90~\GeVcsq. If $\Delta M_1>0$, an
intermediate state exists, and we further require that $\Delta
M_1>$ 5~\GeVcsq~and $(\Delta M_2 - \Delta M_1)>$5~\GeVcsq, i.e., the
intermediate state mass should also be at least 5~\GeVcsq~away from
the upper and the lower states. The results are
accurate to 20-30\% in general and to 30-40\% in regions closer to the extreme edges of the
applicability range.


\begin{table}[htbp]\centering
 \caption{ 
Coefficients in the Taylor expansion of $\{\sigma \! B\}_i^{-1}$ experimental
sensitivities for our generalization 
of the CDF trilepton search (Eqs.~\ref{eq:sigmaB} to \ref{eq:hnew}).
These results are for a 95\% confidence level upper limit with 2~fb$^{-1}$
integrated luminosity and should be scaled if the data amount is
different. 
Coefficient units are such that masses
in GeV/c$^2$ give cross-section in picobarns. These numbers are available 
electronically in a spreadsheet
utility~\cite{pubarch}.
}
 \begin{tabular}{ccccccccc}
 \hline \hline
   && $0~\tau$'s && $1~\tau$'s && $2~\tau$'s && $3~\tau$'s \\ \hline
&   \multicolumn{2}{c}{$f(\rm{M})$} & && && \\
   $a_1$&& $2.70\times 10^{-2}$&& $4.48\times 10^{-2}$&& $5.61\times 10^{-2}$&& $4.27\times 10^{-2}$ \\
   $a_2$&& $-9.48\times 10^{-5}$&& $-1.69\times 10^{-4}$&& $-2.29\times 10^{-4}$&& $-1.59\times 10^{-4}$ \\
&& && && && \\
&   \multicolumn{2}{c}{$g(\Delta M_2)$} & && && \\
   $b_0$&& $-4.39 $&& $-3.59 $&& $-3.71\times 10^{-2}$&& $2.11\times 10^{-1}$ \\
   $b_1$&& $3.28\times 10^{-1}$&&  $1.72\times 10^{-1}$&&  $6.60\times 10^{-4}$&&  $-8.20\times 10^{-3}$ \\
   $b_2$&& $-2.08\times 10^{-3}$&&  $-9.41\times 10^{-4}$&&  $1.51\times 10^{-4}$&&  $1.13\times 10^{-5}$ \\
&& && && && \\
&   \multicolumn{2}{c}{$h(\Delta M_1,\Delta M_2)$} & && && \\
 $c_0$ && $-2.84 $&& $-1.73 $&& $-2.67\times 10^{-1}$&& $1.22\times 10^{-2}$ \\
 $c_1$ && $1.92\times 10^{-1}$&&  $9.66\times 10^{-2}$&&  $8.71\times 10^{-3}$&&  $-9.86\times 10^{-4}$ \\
 $d_1$ && $-3.60\times 10^{-2}$&& $-3.74\times 10^{-2}$&& $-4.36\times 10^{-3}$&& $-1.01\times 10^{-3}$ \\
 $c_2$ && $-1.56\times 10^{-3}$&& $-7.36\times 10^{-4}$&& $-7.91\times 10^{-5}$&&  $8.02\times 10^{-6}$ \\
 $d_2$ && $-2.40\times 10^{-3}$&& $-1.49\times 10^{-3}$&& $-5.25\times 10^{-4}$&& $-1.58\times 10^{-4}$ \\
 $e_2$ && $2.72\times 10^{-3}$&&  $1.73\times 10^{-3}$&&  $5.66\times 10^{-4}$&&  $1.59\times 10^{-4}$ \\
\hline \hline
 \end{tabular}
 \label{tab:par}
\end{table} 

A fairly extensive amount of simulation is required to obtain the 
full parametrization from data as above, and it may be
difficult to carry out the detailed (full) experimental simulation for
each grid point in the parameter space.  A convenient experimental
strategy would then be to use a hybrid simulation scheme consisting of
a fully simulated sparse grid that is filled with a finer grid of points
generated with faster standalone simulation as we do above.

\subsection{Example: A Simple Toy Model}

Consider a supersymmetric scenario with the following mass spectrum :
M($\cone$)=M($\ntwo$)= 150~\GeVcsq, M($\slep$)=130~\GeVcsq, and
M($\none$)(=$M$)=110~\GeVcsq. Suppose that the $\ntwo$'s always contribute
$2\tau$'s to the trilepton state because it decays via
$\stau$'s, but the
$\cone$'s leptonic decays occur democratically to electron, muon and $\tau$ lepton
via the three slepton flavors. This would imply
branching fractions of zero for trileptons with 0 and 1~$\tau$'s ($B_0 = B_1 = 0$)
and that the 2$\tau$'s occur twice as frequently as 3 $\tau$'s. Arbitrarily, we
pick $B_2 = 0.2$ and $B_3=0.1$, implying that 70\% of the decays do not yield trileptons.


The three mass parameters then are
$M_0$(=$M(\none$)) = 110~\GeVcsq, $\Delta M_1 = 20$~\GeVcsq, and $\Delta M_2 = 40$~\GeVcsq. Using
the Taylor expansion results from table~\ref{tab:par}, we get 
the upper limit experimental sensitivies at 95\% confidence 
level for this mass spectrum with 2~fb$^{-1}$ data to be 
$\{\sigma \! B\}_0^{-1}$=8.02~pb$^{-1}$,
$\{\sigma \! B\}_1^{-1}$=3.87~pb$^{-1}$,
$\{\sigma \! B\}_2^{-1}$=0.49~pb$^{-1}$ and
$\{\sigma \! B\}_3^{-1}$=0.11~pb$^{-1}$.
The next step is to fold in the
model's $\tau$ channel branching fractions ($B_0=B_1=0, B_2=0.2$, and $B_3=0.1$) using
equation~\ref{eq:sigmaB} to get the upper limit $\sigma_{XM}$ of 9.2~pb. If the
model's chargino-neutralino production cross section is higher than this value, then the model is
ruled by this CDF result at more than 95\% confidence level.

A spreadsheet utility to carry out the procedure demonstrated in this example is 
available electronically~\cite{pubarch}.

\section{Recovering CDF's mSUGRA Result}
\label{sec:recdf}

\begin{figure} [htb]
  \centering\epsfig{file=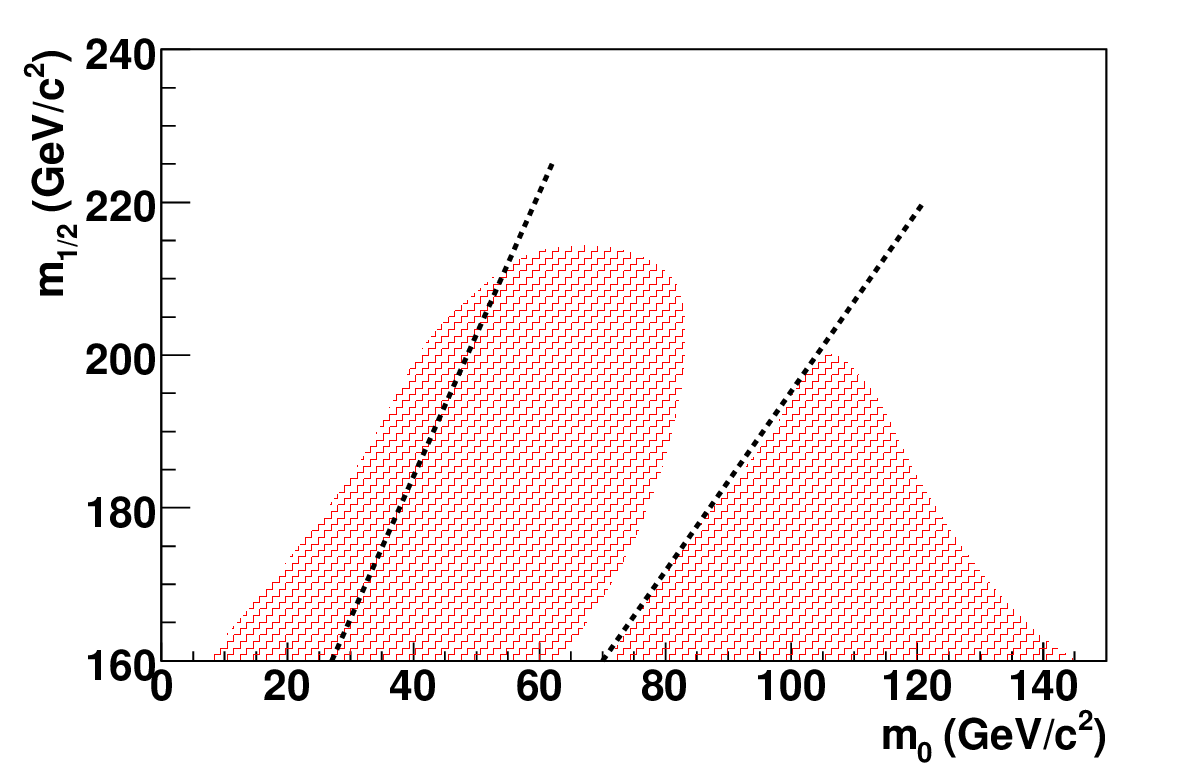,width=10cm}
  \caption{ The mSUGRA exclusion (shaded region) obtained with our generalization of the CDF result. It compares 
favorably with CDF's original exclusion (Figure~2 in Ref.~\cite{cdf2fb}). The dashed line on the right indicates equal chargino
and intermediate slepton masses and similar line on the left indicates equal chargino
and intermediate sneutrino masses.}
  \label{fig:pytexcl2}
\end{figure}

The trilepton search results from the Tevatron have been restricted to the mSUGRA scenario. 
We used the framework described above to generalize the CDF 2~fb$^{-1}$ result.
In order to establish the veracity of our scheme, we apply the parametrized 
formulation above to the very region of mSUGRA parameter space addressed by CDF
(tan($\beta$)=3, $A_0$=0 and $\mu > 0$).  The result
is shown in Figure~\ref{fig:pytexcl2} as two exclusion lobes. The dashed line on the right indicates
where the chargino and the intermediate slepton masses are equal and the one on the left is where
the chargino mass equals that of the intermediate sneutrino. Thus, the $m_0$-$m_{1/2}$ parameter
space shown in the figure is split into three regions: no intermediate state for the rightmost
region, one for the middle region and two for the leftmost region. Our exclusion
curves compare well with CDF's (Figure~2 in Ref.~\cite{cdf2fb}; also see Figure~8 in Ref.~\cite{d02fb}),
but the left exclusion lobe is somewhat smaller than CDF's in the region where
two intermediate particles play a role in the trilepton kinematics. 
Since our parametrization scheme allows at most one intermediate particle between the
chargino and the LSP, we maintain consistency by ignoring the signal 
decays that occur via the sneutrino, thereby lowering our estimate of the sensitivity in that region.

Having verified that our scheme for making experimental results model-independent
works along the expected lines, 
we now apply the Tevatron results to a couple of supersymmetry scenarios
that could not be evaluated given the mSUGRA-specific nature of the published
Tevatron results.

\section{Other Supersymmetry Scenarios and Projections}
\label{sec:otherm}
\begin{figure} [htb]
  \centering\epsfig{file=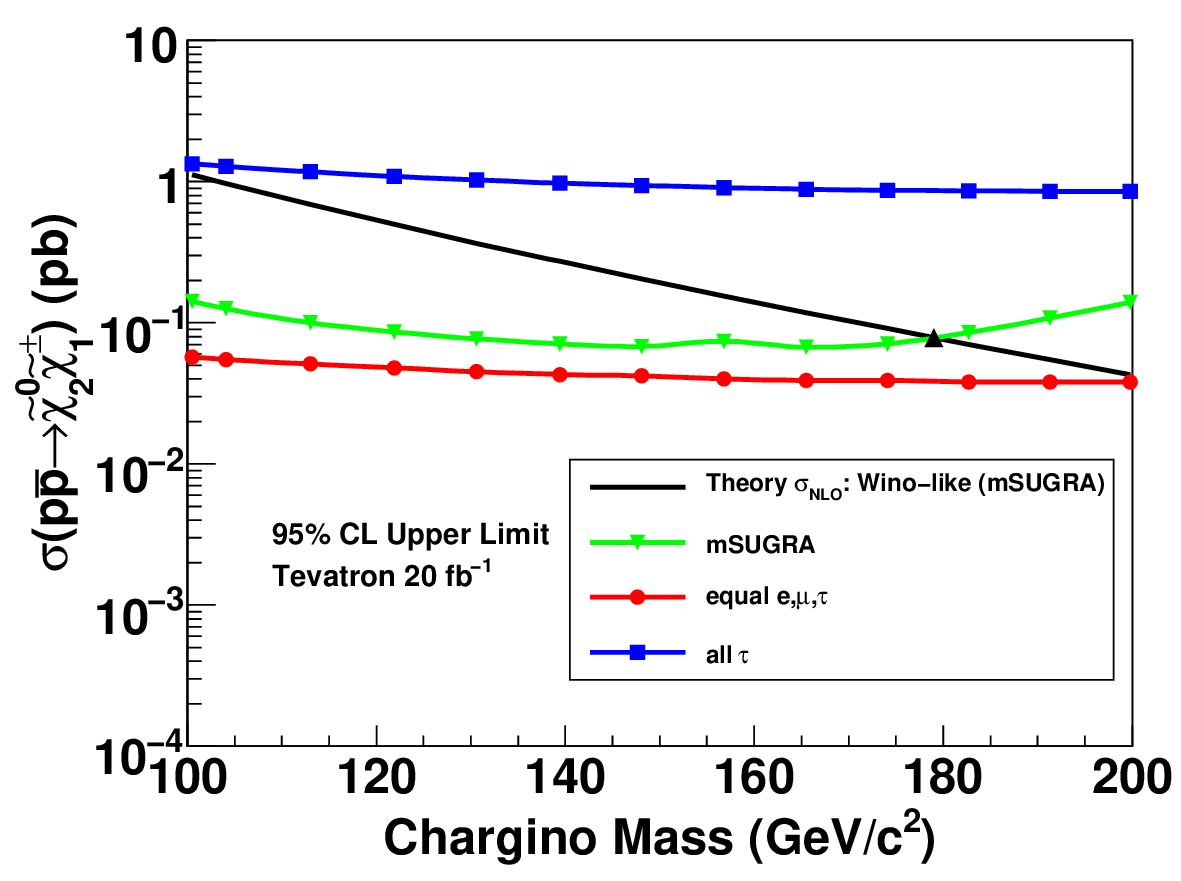,width=10cm}
  \caption{Tevatron trilepton reach for 20~\fbinv for three scenarios 
described in the text.}
  \label{fig:reach}
\end{figure}

We now examine the Tevatron trilepton search reach with 20~fb$^{-1}$ data
for three different supersymmetry scenarios.
The results are shown in Figure~\ref{fig:reach} as 95\% confidence level upper
limit sensitivities, or equivalently, as 1.64~$\sigma$ measurement potentials.

The green curve (inverted triangles) depicts the normal mSUGRA scenario
used by the Tevatron experiments. The chargino mass upper 
limit of approximately 180~GeV/c$^2$ comes from the intersection 
(at the upright triangle) of the green curve with the solid black curve 
representing cross section for wino-like production,
of which mSUGRA is a special case. The
particle masses and branching ratios are all dictated by the parameters of
mSUGRA. For example, the total trilepton branching ratio for the chargino
mass of 150~GeV/c$^2$ is 92.4\%. This curve for experimental sensitivity
is obtained by using our
parametrization for the mSUGRA mass values and then using Equation (\ref{eq:sigmaB})
to blend in the mSUGRA branching ratios. A multiplicative factor takes care of
the ten-fold increase in luminosity over the CDF's.

The figure also shows sensitivity as a function of the upper state (chargino)
mass for two other scenarios for which the mass difference between the upper
state and the intermediate state, $\Delta M_1$,  is held fixed at 30~GeV/c$^2$
and the upper and lower state mass difference,$\Delta M_2$,  at 60~GeV/c$^2$.
Additionally, for the blue curve (squares), all three leptons are forced 
to be $\tau$'s, i.e., $B_0 = B_1 = B_2 = 0$, and $B_3=1$.
As expected, this all-$\tau$ scenario shows a poor
sensitivity due to the experimental difficulty in detecting the $\tau$ lepton.

Finally, for the red curve (circles), all
lepton flavors are treated democratically, giving the trilepton
branching ratio values of
$B_0$ = 4/9, $B_1 = B_2$ = 2/9 and $B_3$ = 1/9. 
The experimental sensitivity for this scenario
is somewhat better than the same for mSUGRA because of a smaller fraction of $\tau$'s.

These scenarios demonstrate how the experimental data presented in our
model-independent scheme can be used with relative ease.
Other models can similarly be confronted by the Tevatron data with
the spreadsheet tool we provide~\cite{pubarch}. 
%
%

\section{Addressing the Inverse Problem}
\label{sec:invprob}

The phrase ``inverse problem'' is used in the context of anticipated
discoveries at the Large Hadron Collider. It refers to the challenge of
identifying the correct theory behind discoveries that
crop up in several experimental channels such as multileptons, photons and
jets.  It could be a difficult problem to solve given the limited
experimental channels in contrast to preponderence of theories and their
vast parametric spaces.

The multichannel cross section equation~\ref{eq:sigmaB}, presented
here in the context of generalizing trilepton supersymmetry search
results, is of a more general use in addressing the ``inverse''
problem of new physics. Let us say that there is simultaneous evidence
for new physics in several channels that have varying degree of
experimental sensitivities for detecting the signal. 
It is quite likely that several theories such as supersymmetry
and technicolor will be put forth as candidates for explaining the
observation.  In addition, these theories have several sub-models and
span a large parameter space. 

Equation~\ref{eq:sigmaB} should be useful for confronting competing
models with the available data.  For each model, the 
sum in equation~\ref{eq:sigmaB} can be
carried out in an extended fashion over all sub-channels of all 
experimental signatures
corresponding to various hypothesized parent states in a theory. The 
grand $\sigma_{XM}$ experimental sensitivity thus obtained can be
compared to the
model's entire cross section for new physics production. 

For example,
one could sum over the $\{\sigma \! B\}_i$ measurements from signatures
such as trilepton, like-sign leptons, diphoton, N-jets, etc., using a
particular SUSY model's branching ratios for chargino-neutralino,
squark-gluino and other experimentally accessible supersymmetric
production processes. The sum can be further extended to include
$\{\sigma \! B\}_i$ measurements from multiple experiments such as CMS and
ATLAS at the LHC. The total $\sigma_{XM}$ thus obtained serves as the
grand experimental measurement of the supersymmetry cross section
predicted by the model under consideration.  The very same underlying
experimental $\{\sigma \! B\}_i$ measurements can be simultaneously used to
confront another model of new physics that may not involve
supersymmetry.

\section{Last Words}
\label{sec:last}

In this paper, we have formulated a recipe for the experimentalists to
present trilepton search results in a model-independent way.  The
inherent problem of a vast parameter space in models of SUSY can be
mitigated by categorizing the experimental sensitivity according to
the $\tau$-lepton content and by expressing it as a function of the
few mass parameters that decide the kinematics of the decay.

Using this method, we showed how to extend the applicability of the
Tevatron trilepton results from their very limited mSUGRA-based
focus. In doing so, we also attempt to bring uniformity to
the disparate methods used by the experiments to interprete
their SUSY trilepton searches. 
Our scheme may also be useful in addressing the broader ``inverse
problem'' of pinpointing new physics if it is discovered in multiple
channels.

\ack{Acknowledgements}
We thank Amit Lath and Matt Strassler of Rutgers University and our CDF collaborators, 
especially Ben Brau, Monica D'Onofrio, 
Chris Hays, Mark Neubauer and Dave Toback. S.T. thanks the Institute for Advanced Study for its hospitality.
The work was supported in part by NSF grant PHY-0650059
and DOE grant DE-FG02-96ER40959. 
The authors are responsible for the contents of this paper and the methodology or interpretation expressed 
in this paper are not endorsed by the CDF collaboration.




\section*{References}
\begin{thebibliography}{9}


\bibitem{SUSYref}
For reviews of the supersymmetric Standard Model see
H.~E.~Haber and G.~L.~Kane,
  ``The Search For Supersymmetry: Probing Physics Beyond The Standard Model,''
  Phys.\ Rept.\  {\bf 117}, 75 (1985);
S.~P.~Martin,
  ``A Supersymmetry Primer,''
  arXiv:hep-ph/9709356;
H.~Baer and X.~Tata,
  ``Weak scale supersymmetry: From superfields to scattering events,''
{\it  Cambridge, UK: Univ. Pr. (2006) 537 p};
 M.~Drees, R.~Godbole, and P.~Roy,
  ``Theory and phenomenology of sparticles: An account of four-dimensional N=1
  supersymmetry in high energy physics,''
{\it  Hackensack, USA: World Scientific (2004) 555 p}.

\bibitem{oldtrilepton}  P.~Nath and R.~L.~Arnowitt,
   ``Supersymmetric Signals at the Tevatron,''
   Mod.\ Phys.\ Lett.\  A {\bf 2}, 331 (1987);
R.~Barbieri, F.~Caravaglios, M.~Frigeni, and M.~L.~Mangano,
   ``Production And Leptonic Decays Of Charginos And Neutralinos In Hadronic
   Collisions,''
   Nucl.\ Phys.\  B {\bf 367}, 28 (1991);
H.~Baer and X.~Tata,
   ``Probing charginos and neutralinos beyond the reach of LEP at the tevatron
   collider,''
   Phys.\ Rev.\  D {\bf 47}, 2739 (1993);
H.~Baer, C.~Kao, and X.~Tata,
   ``Aspects of chargino - neutralino production at the tevatron collider,''
   Phys.\ Rev.\  D {\bf 48}, 5175 (1993)
   [arXiv:hep-ph/9307347].



\bibitem{cdf2fb} T. Aaltonen \textit{et al.} (CDF Collaboration), 
Phys. Rev. Lett.~\textbf{101}, 251801 (2008).

\bibitem{d0trilep} V.~Abazov \textit{et al.} (\Dzero Collaboration), 
Phys. Rev. Lett.~\textbf{95}, 151805 (2005).

\bibitem{d02fb} V.~Abazov \textit{et al.} (\Dzero Collaboration), 
Phys. Lett. B~\textbf{680}, p34-43 (2009).


\bibitem{pythia}
T.~Sj\"{o}strand, P.~Eden, C.~Friberg, L.~Lonnblad, G.~Miu, S.~Mrenna, and E.~Norrbin, Comput. Phys. Commmun. \textbf{135}, 238 (2001) (we use version 6.216).

\bibitem{juls} J.~Glatzer, ``Probing mSUGRA with a search for chargino-neutralino production using trileptons'', FERMILAB-MASTERS-2008-05, Appendix D.

\bibitem{thesis} S.~S.~Dube, ``Search for supersymmetry at the Tevatron using the trilepton signature,'' FERMILAB-THESIS-2008-45.


\bibitem{pubarch} Parametrizations included in this article and other information
is available electronically from Rutgers Physics Publications Archive at
http://www.physics.rutgers.edu/pub-archive/0901

\bibitem{root} R.~Brun and F.~Rademakers,``ROOT - An Object Oriented Data Analysis Framework'',
Proceedings AIHENP'96 Workshop, Lausanne, Sep. 1996, Nucl. Inst. \& Meth. in Phys. Res. A 389 (1997) 81-86. See also http://root.cern.ch/

\end{thebibliography}
\end{document}